\documentstyle[preprint,aps]{revtex}

\def\xandy{\mathop{|x\rangle |y\rangle}}
\def\xands{\mathop{|x\rangle |s\rangle}}
\def\sands{\mathop{|s\rangle |s\rangle}}

\def\psands{\mathop{|\psi\rangle |s\rangle}}

\def\1{{1\kern-.25em\hbox{\rm I}}}
\def\eu{{1\kern-.25em\hbox{\sm I}}}

\begin{document}

\title{Quantum Mechanical Square Root Speedup in a\\
Structured Search Problem\thanks{\baselineskip=16pt This work was supported in
part by The Department of Energy under cooperative agreement DE-FC02-94ER40818}}

\def\footstrut{\baselineskip 16pt}

\author{\large Edward Farhi}
\address{Center for Theoretical Physics \\ 
Massachusetts Institute of Technology \\
Cambridge, MA  02139}

\author{\large Sam Gutmann}
\address{Department of Mathematics \\ 
Northeastern University \\
Boston, MA  02115\\ [2em]
{\rm MIT-CTP-2691,~~quant-ph/9711035 \hfil \qquad \qquad November 1997}}

\maketitle

\tightenlines

\begin{abstract}%
An unstructured search for one item out of $N$ can be performed quantum
mechanically in time of order $\sqrt{N}$ whereas  classically this requires of order
$N$ steps.  This raises the question of whether
square root speedup persists in problems with more structure. In this note we focus
on one example of a structured problem and find a quantum algorithm which takes
time of order  the square root of the classical time.  
\end{abstract}

\newpage

\section*{Introduction}

An unstructured search for one item out of $N$ can be performed quantum
mechanically in time of order $\sqrt{N}$ whereas  classically this requires of order
$N$ steps \cite{Grov}.  The $\sqrt{N}$ is optimal\cite{Prosh}.  This raises the question
of whether square root speedup persists in problems with more structure \cite{ref:3}.
In this note we focus on one example of a structured problem and find an (optimal)
quantum algorithm which takes time of order  the square root of the classical time. 
  Some of the
methods in this paper are similar to those found in\cite{ref:4}.

\section{Main Result}

Consider a function $F(x,y)$ with $x$ and $y$ integers, $1 \le x \le L$ and $1\le y \le
L$.  The function has the property that it is 0 except at a single value of $(x,y)$ where
it takes the value 1.  We imagine that a subroutine which computes $F$ is available
but  we have no further knowledge of $F$.  The goal is to discover the unique $(x_0,
y_0)$ where $F(x_0, y_0)=1$.  As described so far, classically it is necessary to search
the $L^2$ values of $(x,y)$ whereas quantum mechanically, Grover's algorithm finds
$(x_0,y_0)$ with of order
$L$ calls of the (quantum) subroutine for $F$.

Now suppose we also have available a subroutine which computes a function $G(x)$,
$1\le x \le L$.  This function $G$ is known to have the property that it takes the value
1 on a set with $M$ elements and is 0 otherwise.  Furthermore we are guaranteed that
$G(x_0)=1$.  The goal is
to find $(x_0,y_0)$ as fast as possible where time is measured in the total number of
function calls of $F$ and $G$.  

If $M$ equals $L$, the function $G$ is identically 1 and
is useless.  If $M$ equals 1, the best strategy --- classically or quantum mechanically
--- is to use $G$ to find $x_0$ and then $F$ to find $y_o$.  We thus restrict our
attention to the case where $1 \ll M\ll L$.  Furthermore we assume that $M$ is known.

Classically, $(x_0,y_0)$ can be found in of order $ML$ steps.  The strategy is to
consider each $x$ in turn.  If $G(x) =0$ move to the next $x$; otherwise check $F(x,y)$
for each $y$.   This time is optimal: the {\bf easier} problem in which we are {\bf told}
that $G(x)=1$, $1\le x \le M$, (and $G(x_0)=1$) already requires a search through
$ML$ items.  The same argument shows that no quantum algorithm can succeed in
fewer than $\sqrt{ML}$ steps.

We now present a quantum algorithm that succeeds in time of order $\sqrt{ML}$.  We
work in an $L^2$ dimensional Hilbert space with orthonormal basis elements
$\xandy$.  Quantum code for $F$ and $G$ allows us to readily construct the
unitary operators $(-1)^{\hat{F}}$ and  $(-1)^{\hat{G}}$ defined by 
\begin{eqnarray}
 (-1)^{\hat{F}} \xandy &=&  (-1)^{F(x,y)} \xandy \label{eq:1} \\
\noalign{\hbox{and}}
 (-1)^{\hat{G}} \xandy &=&  (-1)^{G(x)} \xandy  \ \ . 
\label{eq:2}
\end{eqnarray}
Furthermore with $|s\rangle$ defined as 
\begin{equation}
|s\rangle = \frac{1}{\sqrt{L}} \sum^L_{z=1} |z\rangle
\label{eq:3}
\end{equation}
we can also construct $\hat{U}_1$ and $\hat{U}_2$ given by
\begin{eqnarray}
\hat{U}_1  &=&  (2|s \rangle \langle s|-1) \otimes 1
\label{eq:4}
\\
\noalign{\hbox{and}}
\hat{U}_2 &=& 1 \otimes (2|s \rangle \langle s|-1)  \ \ . 
\label{eq:5}
\end{eqnarray}
All of these are immediate generalizations of the building blocks of Grover's algorithm
\cite{Grov}.

Now consider the operator $\hat{W}$ defined by 
\begin{equation}
\hat{W}= [\hat{U}_2 (-1)^{\hat{F}}] ^k
\label{eq:6}
\end{equation}
where $k$ is the closest integer to $\frac{\pi}{4}\sqrt{L}$.  What is $\hat{W} \xands$? 
If $x\ne x_0$, then $F(x,y)=0$ for all $y$ and $\hat{W}$ acts as the identity. If
$x=x_0$, then $\hat{W}$ is executing the Grover algorithm on the $y$ coordinate with
the function $F(x_0, y)$ and $\hat{W}|x_0\rangle|s\rangle= |x_0\rangle |y_0\rangle$.
(The algorithm actually produces a state which is $|x_0\rangle |y_0\rangle$ plus
corrections of order $\frac{1}{\sqrt{L}}$.  Throughout this paper we ignore these
corrections.) Thus
\begin{equation}
\hat{W}\xands = \left\{%
\begin{array}{lcl}
\xands &{\rm if} & x \ne x_0 \\
|x_0\rangle|y_0\rangle &{\rm if} & x = x_0  \ \ .
\end{array}
\right. 
\label{eq:7}
\end{equation}
Next consider
\begin{equation}
\hat{V}= \hat{W}^\dagger \,\, (-1)^{\hat{F}}  \,\, \hat{W} \ \ .
\label{eq:8}
\end{equation}
It follows from (\ref{eq:7}) that
\begin{equation}
\hat{V}\xands = \left\{%
\begin{array}{lcl}
|x\rangle|s\rangle &{\rm if} & x \ne x_0 \\
-|x_0\rangle|s\rangle &{\rm if} & x = x_0  \ \ .
\end{array}
\right. 
\label{eq:9}
\end{equation}
We will use $\hat{V}$ as a subroutine in our overall algorithm.

We now take advantage of the function $G$.  First define the superposition of
$|x\rangle$'s for which $G(x)=1$,
\begin{equation}
|\psi\rangle = \frac{1}{\sqrt{M}} \sum_{G(x)=1} |x\rangle \ \ .
\label{eq:10}
\end{equation}
Note that $\psands$ can be obtained from $\sands$  by a
straightforward generalization\cite{ref:5} of the Grover algorithm to the case where
the number of marked items is $M$,
\begin{equation}
 [\hat{U}_1 (-1)^{\hat{G}}]^j \sands = \psands
\label{eq:11}
\end{equation}
where $j$ is the closest integer to $\frac{\pi}{4} \sqrt{\frac{L}{M}}$. (The $y$
coordinate is just coming along for the ride.)  Furthermore we can construct
\begin{equation}
\hat{U}_\psi = (2|\psi \rangle \langle \psi | -1) \otimes 1
\label{eq:12}
\end{equation}
by
\begin{equation}
\hat{U}_\psi = [ (-1)^{\hat{G}} \hat{U}_1]^j \,\, \hat{U}_1 \,\, [\hat{U}_1
(-1)^{\hat{G}}]^j \ \ .
\label{eq:13}
\end{equation}
Using $\hat{U}_\psi$ and $\hat{V}$ we can use the Grover algorithm to find $x_0$
from a set of $M$ things instead of from a set of $L$ things.  That is, with $h$ being the
closest integer to $\frac{\pi}{4} \sqrt{M}$, we have that
\begin{equation}
[\hat{U}_\psi\hat{V}]^h \psands = |x_0\rangle  |s\rangle \  \ . 
\label{eq:14}
\end{equation}

To run the algorithm we start in the state $|s\rangle |s\rangle$, apply (\ref{eq:11})
and then (\ref{eq:14}) obtaining $|x_0\rangle |s\rangle$.  How many steps does it take
to produce $|x_0\rangle |s\rangle$ by this method?  Up to constants, it takes
$\sqrt{L}$ calls of $F$ to construct $\hat{W}$ (see (\ref{eq:6})) and accordingly
$\sqrt{L}$ calls of $F$ to construct $\hat{V}$ (see (\ref{eq:8})).  Now
$\hat{U}_\psi$ takes $\sqrt{\frac{L}{M}}$ calls of $G$ (see (\ref{eq:13})), so
$\hat{U}_\psi\hat{V}$ in (\ref{eq:14}) takes $\sqrt{\frac{L}{M}} + \sqrt{L}$ which is
$\sqrt{L}$.  Thus producing $|x_0\rangle |s\rangle$ from $|\psi\rangle |s\rangle$ takes
$h\sqrt{L}$, that is $\sqrt{ML}$ steps.  (Note that producing $|\psi\rangle |s\rangle$ 
by (\ref{eq:11}) requires an irrelevant extra $\sqrt{\frac{L}{M}}$ steps.)

We have found $x_0$ in of order $\sqrt{ML}$ steps.  Using the function $F(x_0,y)$ we
can quantum mechanically search through the $L$ values of $y$ in time $\sqrt{L}$. 
The total time required to find $(x_0,y_0)$ remains of order $\sqrt{ML}$.  

\bigskip

\section{Another Example}

\medskip

The problem just discussed is a more structured version of the following.  We are
given a subroutine for a function $f(z)$ with $1 \le z \le N$ which is guaranteed to be 0
except at a unique but unknown point $z_0$.  Suppose we are also given a function
$g(z)$ and we know that $g(z)=1$ for $z$ in a set of size $M$ and that $z_0$ is in this
set.  A natural way to try to use $g$ to speed up the quantum search for $z_0$ is to (i)
use the Grover algorithm to construct the state $|\phi \rangle$ which is a superposition
of the $M$ basis states $|z\rangle$ with $g(z)=1$ and then (ii) use the Grover algorithm
on $|\phi \rangle$ with the function $f$.  We now explicitly do this and find that
the total time is of order $\sqrt{N}$ so that the added ability to call
$g$ is of no help.

To begin we define
\begin{equation}
|\sigma\rangle = \frac{1}{\sqrt{N}} \sum^N_{z=1} |z\rangle
\label{eq:15}
\end{equation}
and
\begin{equation}
|\phi \rangle = \frac{1}{\sqrt{M}} \sum_{g(z)=1} |z\rangle \ \ .
\label{eq:16}
\end{equation}
The state $|\phi \rangle$ can be obtained, as in (\ref{eq:11}), with $\ell$ applications
of 
\begin{equation}
\hat{U} = [2|\sigma \rangle \langle \sigma|-1] (-1)^{\hat{g}}
\label{eq:17}
\end{equation}
where $\ell$ is the integer closest to $\frac{\pi}{4}
\sqrt{\frac{N}{M}}$; that is
\begin{equation}
\hat{U}^\ell \,\, |\sigma\rangle = |\phi \rangle \ \ .
\label{eq:18}
\end{equation}
As in (\ref{eq:12}) and (\ref{eq:13}), we can produce
\begin{equation}
2|\phi \rangle \langle \phi|-1  = \hat{U}^\ell \,\, (2|\sigma \rangle \langle
\sigma|-1) \,\,
\hat{U}^\ell \ \ .
\label{eq:19}
\end{equation}
Finally
\begin{equation}
[(2|\phi \rangle \langle \phi|-1) (-1)^{\hat{f}}]^h \,\, |\phi\rangle = |z_0 \rangle
\label{eq:20}
\end{equation}
where $h$ is the closest integer to $\frac{\pi}{4} \sqrt{M}$.

How many function calls does it take to produce $|z_0 \rangle$?  Since 
$2|\phi \rangle \langle \phi|-1$ requires of order $\sqrt{\frac{N}{M}}$ calls of $g$ and
we must use $2|\phi \rangle \langle \phi|-1$ of order $\sqrt{M}$ times, we need of
order $\sqrt{N}$ calls of $g$.  Thus the availability of $g$ yields no speedup over the
Grover algorithm using $f$ alone.

It is easy to see why using $g$ is never of computational benefit --- quantum
mechanically or classically.  Suppose it was of benefit and an algorithm using $f$ and
$g$ existed which was {\bf faster} than any algorithm using $f$ alone.  Now given only
a function $f(z)$ which is 1 at a unique unknown $z_0$, one could always construct two
functions $g_1(z)$ and $g_2 (z)$ as
\begin{equation}
g_1(z) = \left\{%
\begin{array}{rl}
1 & \quad {\rm if~} 1 \le z \le M-1 \quad {\rm or} \quad f(z)=1\\
0 & \quad {\rm otherwise}
\end{array}
\right. 
\label{eq:21}
\end{equation}
and
\begin{equation}
g_2(z) = \left\{%
\begin{array}{rl}
1 & \quad {\rm if~} 1 \le z \le M \quad {\rm or} \quad f(z)=1\\
0 & \quad {\rm otherwise} \ \ .
\end{array}
\right. 
\label{eq:22}
\end{equation}
Note that if $1 \le z_0 \le M$ then $g_2$ is 1 for $M$ values of $z$ (including $z_0$)
whereas if $M \le z_0 \le L$ then $g_1$ is 1 for $M$ values of $z$ (including $z_0$). 
Run the purportedly faster algorithm using first $f$ and $g_1$ and then with $f$ and
$g_2$.  This would produce $z_0$ as quickly as the faster algorithm, which is a
contradiction.

The example in this section shows that whenever the Grover algorithm is used within
the Grover algorithm\cite{ref:4}, the time required to construct the operator $2|\phi
\rangle \langle \phi|-1$ must be taken into account.

\section{Discussion}

In the first section we gave an example of a structured search problem where the
(best possible) quantum algorithm succeeded in a time of order the square root of the
classical time.  The example in Section 2 also has this feature.  Here the apparent
structure introduced through the function $g$ does not allow the classical search to
be done in time faster than $N$ nor does it allow the quantum search to be done faster
than $\sqrt{N}$.  These two examples give evidence that quantum square root speed
up may persist in a wide range of structured problems.

\vskip.2in

\noindent{\bf Acknowledgments}

We would like to thank Jeffrey Goldstone and Norm Margolus for helpful discussions. 
We also thank Lov Grover for sending us an early draft of \cite{ref:4} and for a
stimulating talk and conversation.


\begin{thebibliography}{99}
\bibitem{Grov} L.~K.~Grover, A fast quantum mechanical algorithm for
database search,
quant-ph/9605043.

\bibitem{Prosh} C.~H.~Bennett, E.~Bernstein, G.~Brassard, and
U.~V.~Vazirani, Strengths
and weaknesses of quantum computing, quant-ph/9701001.


\bibitem{ref:3} J. Preskill, Quantum Computing: Pro and Con, quant-ph/9705032.

\bibitem{ref:4} L. Grover, Any unitary system can perform rapid search.

\bibitem{ref:5} M.~Boyer, G.~Brassard, P.~Hoeyer, and A.~Tapp, Tight bounds on
quantum searching, quant-ph/9605034.

\end{thebibliography}
\end{document}